\documentclass[aps,pre,twocolumn,groupedaddress]{revtex4-1}
\usepackage{amsmath,amssymb,graphicx,dsfont,tikz,subfigure}
\usepackage[english]{babel}
\usepackage[utf8]{inputenc}
\usepackage{stmaryrd}
\usepackage{epstopdf}

\newcommand{\vq}{\mathbf{q}}
\newcommand{\vep}{\mathbf{p}}
\newcommand{\vp}{\mathbf{p}}
\newcommand{\vf}{\mathbf{f}}
\newcommand{\vg}{\mathbf{g}}
\newcommand{\bvq}{\bar{\mathbf{q}}}
\newcommand{\bq}{\bar{q}}
\newcommand{\hq}{\hat{q}}
\newcommand{\bu}{\bar{u}}
\newcommand{\vu}{\mathbf{u}}
\newcommand{\by}{\bold{y}}
\newcommand{\pa}{\partial}
\newcommand{\eps}{\varepsilon}
\newcommand{\al}{\alpha}
\newcommand{\la}{\lambda}

\newcommand{\lan}{\langle}
\newcommand{\ran}{\rangle}

\begin{document}
\title{Large deviations in Taylor dispersion}

\author{Marcel Kahlen}
\author{Andreas Engel}
\author{Christian Van den Broeck}
\affiliation{Universit\"at Oldenburg, Institut f\"ur Physik, 26111 Oldenburg, Germany \\ Hasselt University, Faculty of Sciences, B-3590 Diepenbeek, Belgium}

\begin{abstract}
We establish a link between the phenomenon of Taylor dispersion and the theory of empirical distributions. Using this connection, we derive, upon applying the theory of large deviations, an alternative and much more precise description of the long-time regime for Taylor dispersion.
\end{abstract}

\pacs{}
\maketitle

\section{Introduction}\label{sec:intro}
Around 1953, G. I. Taylor  \cite{Taylorrev} published a series of papers in which he studies the dispersion of particles injected in a Poiseuille flow through a cylindrical tube.  As the particles are carried along, their elution profile broadens along the flow axis $x$ into a co-moving Gaussian shape, with a mean square displacement described by an effective diffusive law,  $\langle [x (t)- \langle x (t) \rangle ]^2 \rangle=2 Kt $.  He  calculated the value of the  "Taylor dispersion coefficient":  $K=\bar{u}^2 a^2/(48 D)$, a result also obtained by Westhaver in his study of electro-migration \cite{Westhaver}. Here $D$ is the molecular diffusion coefficient of the suspended particles, $\bar{u}$ the average flow velocity and $a$ the radius of the tube. As an application, he discusses how such a set-up allows to estimate $D$, which is typically very small, by measuring $K$, a method used routinely in biophysics by now \cite{comapp}. The catch is that the Gaussian regime is attained for times  larger than the typically very long time $a^2/D$ for the particles to diffuse over the tube's radius.  The dispersion of particles in a flow is one example of a more general phenomenon, referred to as generalized Taylor dispersion, which we  briefly review below, and which includes applications in chromatography, the sedimentation of non-spherical particles, spin relaxation and phase diffusion in limit cycles, models for kinetic theory, random evolution, etc. \cite{Broeckrev,brenner}.

After pioneering work by H.~Cramer and I.~N.~Sanov, the mathematical theory of large deviations was primarily shaped in a series of influential papers by M.~D.~Donsker and S.~R.~S.~Varadhan \cite{Donskerrev1}. Recent years have seen an ever increasing number of studies of large deviations in statistical mechanics, see, e.g., \cite{Touchetterev,Derrida,Vulpiani}. The Gaussian distribution as it appears in Taylor's problem corresponds to the quadratic approximation for the so-called large deviation function (ldf). The explicit evaluation of the ldf for Taylor dispersion was done in \cite{HaVa} and provides a more precise description of the large time regime. But, as we will show in the present paper, the connection between both problems runs deeper: the large deviation description for generalized Taylor dispersion can be mapped onto that for empirical distributions. The empirical distribution is the probability distribution observed in a sample realization of a stochastic process. It plays a crucial role in statistics and its  properties form a  cornerstone of  large deviation theory proper. The connection between Taylor dispersion and empirical distributions allows to compare and exchange techniques and results between both fields of research. 

The paper is organized as follows. In section \ref{sec:gen} we shortly review generalized Taylor dispersion and explicitly establish the relation to the theory of empirical distributions. We then turn in section \ref{sec:disc} to the large deviation analysis of Taylor dispersion in layered systems for which the underlying stochastic process acts in a discrete state space. In section \ref{sec:cont} we discuss continuous flow fields including the Poisieulle flow studied originally by Taylor. Finally, section \ref{sec:discussion} contains some concluding remarks. 


\section{General theory}\label{sec:gen}
\begin{figure}[h]
  \includegraphics[trim = 0cm 0cm 0cm 0cm, clip, width=8cm]{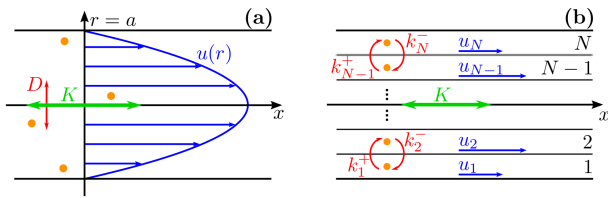}\vspace{-0.1cm}
  \caption{a) Taylor dispersion for a parabolic flow profile $u(\bold{y})$. Molecular diffusion with coefficient $D$ perpendicular to the flow gives rise to a dispersion of particles parallel to the flow with effective diffusion constant $K$. b) In a discretized model the particles reside in $N$ layers with drift velocities $u_i$ and make random transitions between adjacent layers with rates $k_i^+$ and $k_i^-$.}\label{fig:Comparison}
\end{figure}

To set the scene, we start with a brief review of generalized Taylor dispersion. A  variable $x$ has a rate of change $u$, which is driven by a  stochastic process $\bold{y}$, independent of $x$: 
\begin{equation}\label{eqm}
\dot{x}(t)=u(\bold{y}(t)).
\end{equation}
In the example described in the introduction, $x$ is the coordinate measured along the tube,  $\bold{y}=\bold{r}$ is the radial position of the particle undergoing molecular diffusion, and 
\begin{equation}\label{uofr}
 u(\bold{r})=2\bar{u}(1-\frac{\bold{r}^2}{a^2})
\end{equation} 
the flow profile, cf. Fig.~\ref{fig:Comparison}a. By integrating the equation of motion, one finds the relation between the sample position $x=x(t)$ at time $t$, and the particular realisation $\bold{y}(\tau), \tau\in [0,t]$, of the stochastic process:
\begin{equation}\label{em}
x(t)=\int_0^t u(\bold{y}(\tau)) d\tau.
\end{equation}
We will assume that  the stochastic process $\bold{y}$ is stationary, with stationary distribution $p(\bold{y})$, and does not display long-time correlations. The resulting displacement $x$ will, for long times, be the sum of essentially uncorrelated and identically distributed contributions, corresponding to subsequent displacements during time intervals larger than the correlation time of $\bold{y}$.  Hence one expects from the central limit theorem that the  distribution for $x$  approaches a Gaussian in the long time limit:
\begin{equation}\label{clt}
P(x,t)\sim\frac{1}{\sqrt{4\pi Kt}} \exp\left\{-\frac{(x-\bar{u}t)^2}{4Kt}\right\},
\end{equation}
with the average speed
\begin{equation}
\langle u(\bold{y}) \rangle=\int u(\bold{y}) p(\bold{y}) d\bold{y}=\bar{u},
\end{equation}
and the Taylor dispersion coefficient given by a Green-Kubo formula:
\begin{equation}\label{K}
K=\int\limits_{0}^{\infty} \langle [u(\bold{y}(\tau))-\bar{u}] [u(\bold{y}(0))-\bar{u}] \rangle d\tau.
\end{equation}

The large deviation analysis of generalized Taylor dispersion builds on the well studied large deviation properties of the empirical density of the underlying stochastic process \cite{Donskerrev1,Shorack}. To make the connection explicit let us consider the case in which the rate of change $u_i$ of the coordinate $x$ depends on a discrete variable $i\in \bold{I}$, rather then the continuous variable $\bold{y}$ discussed before, cf. Fig.~\ref{fig:Comparison}b. The equation of motion (\ref{eqm}), is thus replaced by $\dot{x}(t)=u_{i(t)}$. The discrete state $i$ undergoes an independent stationary stochastic process with steady state distribution $p_i$. 
During a realization of the stochastic process of duration $t$, the system will spend relative times 
\begin{equation}\label{defq}
 q_i(t)=\frac{1}{t}\int_0^t \delta_{i,i(\tau)} d\tau
\end{equation} 
in each of the states $i\in \bold{I}$. The object, $\bold{q}=\{q_i, i\in \bold{I}\}$, called the empirical density and  itself a random variable, is to be distinguished from the "genuine" distribution $\bold{p}=\{p_i, i\in \bold{I}\}$. Nevertheless, for an ergodic process $\bold{q}$ will converge to $\bold{p}$ in the long-time limit. More precisely, one finds for large $t$ 
\begin{eqnarray}\label{ldtl2}
P(\bold{q},t)\sim \exp\{-t J(\bold{q})\},
\end{eqnarray}
with the ldf $J(\bold{q})$ satisfying $J(\bold{p})=0$ and $J(\bold{q})>0$ for $\bold{q}\neq \bold{p}$. In other words,  deviations of $\bold{q}$ from $\bold{p}$ are exponentially suppressed in the long-time limit. 

Instead of considering the sample displacement  at time $t$, $x=x(t)$, we now focus on the sample velocity $v=v(t)=x(t)/t$, which is clearly given by:
\begin{eqnarray}\label{ed}
v=\sum_i  u_i q_i=\bold{u\cdot q}.
\end{eqnarray}
This expression links the stochastic properties of $v$ to those of the empirical distribution $\bold{q}$. In particular, the large time properties of the sample speed are specified by a large deviation principle of the form
\begin{eqnarray}\label{ldt}
P(v,t)\sim \exp\{-t I(v)\},
\end{eqnarray}
with the ldf $I(v)$ linked to $J(\bold{q})$ by the contraction principle of large deviation theory \cite{Touchetterev}:
\begin{equation}\label{IJ}
I(v)=\inf_{\bold{u\cdot q}=v \;\bold{1\cdot q}=1 } J(\bold{q}).
\end{equation}
This property corresponds to the intuitively obvious observation that an exponentially rare sample speed $v$ is realised by the least unlikely, exponentially rare realisation of an empirical distribution $\bold{q}$. We have explicitly included the constraint that $\bold{q}$ is normalized, $\sum_i q_i=\bold{1\cdot q}=1$.
Note that the central limit approximation, cf. Eq. (\ref{clt}), corresponds to the  quadratic approximation of the ldf around its minimum, 
\begin{equation}\label{cltI}
I(v)\cong\frac{(v-\bar{u})^2}{4K},
\end{equation}
whereas the exact large deviation function will have contributions of higher order in $v$. 

To proceed, we will assume that the stochastic process $i$ is Markovian, and represents an equilibrium situation. This covers most applications in generalized Taylor dispersion. The Markov process is characterized by a transition matrix $\bf{W}$. Its elements are the rates $W_{i,j}$, corresponding to the probability per unit time to make a transition from state $j$ to state $i$ for $i \neq j$, and  ${W}_{i,i}=-\sum_{j,j\neq i} W_{j,i}$.  They obey the detailed balance property:
\begin{eqnarray}\label{db}
W_{j,i}p_i=W_{i,j}p_j.
\end{eqnarray}
In this case, the ldf of the empirical distribution is known explicitly \cite{Donskerrev1}:
\begin{eqnarray}\label{lded}
J(\bold{q})=\frac{1}{2}\sum_{i,j}(\sqrt{{W}_{i,j}q_j}-\sqrt{W_{j,i}q_i})^2.
\end{eqnarray}
To account for the constraints in the contraction \eqref{IJ} we introduce two Lagrange multipliers $\lambda_1$ and $\lambda_2$. The vector $\bar{\bold{q}}$ minimizing 
\begin{equation}\label{eq:barq}
 J(\bold{q})+\lambda_1(\bold{1}\cdot\vq -1)+\lambda_2(\vu\cdot\vq -v)
\end{equation} 
has to fulfill the system of linear equations: 
\begin{equation}\label{eq:contr}
 (\lambda_1+\lambda_2 u_i)\sqrt{\bar{q}_i}-\sum_j\sqrt{W_{i,j}W_{j,i}}\sqrt{\bar{q}_j}=0\;\;\;\forall i\in \bold{I}.
\end{equation} 
Multiplying by $\sqrt{\bar{q}_i}$ and summing over $i$ yields:
\begin{equation}\label{Isimple}
 I(v)=J(\bar{\bold{q}})=-(\lambda_1+\lambda_2 v).
\end{equation} 
So all that needs to be determined are the values of the Lagrange multipliers $\lambda_1$ and $\lambda_2$. 
In general there will be several solutions. To select the proper one it is instructive to write \eqref{eq:contr} in matrix form:
\begin{align}\label{eq:N-layer model Sturm Liouville}
    \bold{L}	\begin{pmatrix}
		\sqrt{\bar{q}_1} \\
		\vdots \\
		\sqrt{\bar{q}_N} 
	\end{pmatrix} = \lambda_1 \begin{pmatrix}
		\sqrt{\bar{q}_1} \\
		\vdots \\
		\sqrt{\bar{q}_N} 
	\end{pmatrix} \; ,
\end{align}
showing that $\lambda_1$ is an eigenvalue of  the symmetric $N\times N$ matrix $ \bold{L}$ with elements  $L_{i,j}=\sqrt{W_{i,j}W_{j,i}}-\lambda_2 u_i \delta_{i,j}$, with all entries of the corresponding eigenvector positive. Hence, by the Frobenius-Perron theorem, the largest eigenvalue of $L$ is the relevant solution for the Lagrange multiplier $\lambda_1$.

As a first application of the above formalism and to demonstrate its consistency with previous work, we rederive the exact expression and bounds for the Taylor dispersion coefficient $K$. In order to capture the Gaussian regime we only need the quadratic expansion of the ldf $I(v)$ around the most probable sample value at $v=\bar{u}$, cf. Eq. (\ref{cltI}). Starting from \eqref{lded} we derive in appendix \ref{sec:A} the general expression \cite{BroeckMazo,Bouten}
\begin{equation}
  K=-\sum_{i,j} \tilde{W}^{-1}_{i,j}\, p_j\,(\bu-u_i)(\bu-u_j).
\end{equation} 
Here $\tilde{W}^{-1}_{i,j}$ denotes the inverse of $W_{i,j}$ on the subspace of non-zero eigenvalues. It is usually difficult if not impossible to evaluate this inverse exactly. As an alternative, the variational form of \eqref{IJ}  invites for the derivation of bounds on $K$. Since $v=\bu$  corresponds to $q_i=p_i$ it suffices to keep the first order deviations $q_i\cong p_i+f_i(v-\bar{u})$ in (\ref{lded}) before doing the contraction. In this way we show in appendix \ref{sec:A} 
\begin{equation}
 K\geq-\frac{1}{\lan \vf|W|\vf\ran},
\end{equation}
where we used the scalar product $\lan \vf|\vg\ran=\sum_i f_i g_i/p_i$ and $\vf$ is any vector perpendicular to $\vp$ and fulfilling $\lan \vf|\vu\vp\ran=1$. This expression is identical to the lower bound for $K$ derived previously in \cite{Bouten}.



\section{Discrete flow fields}\label{sec:disc}
Turning to the large deviation properties, we first focus on the simplest model capturing the essential ingredients  of Taylor dispersion. It consists of just two layers with velocities $u_1$ and $u_2$, respectively. Because of Galilei invariance we may choose $u_2=-u_1=:u$ without loss of generality. We denote by  $k_+=W_{2,1}$ and  $k_-=W_{1,2}$ the rates of transition from layer $1$ to $2$ and back, respectively. In this case, the exact elution profile is known for all times, and was derived in many different settings \cite{Goldstein,Giddingsrev,Mysels,McKean,Kac,Thacker}.
\begin{eqnarray}
   P(x, t)= && \frac{1}{4\pi}\left[\frac{\partial}{\partial t}\! + \!(k_+\!+ k_-)\right]  \nonumber\\ 
   &&\left[\frac{2\pi}{u}\exp{\left(-\frac{k_+}{2}(t-\frac{x}{u})-\frac{k_-}{2}(t+\frac{x}{u})\right)}\right. \nonumber \\ 
 && \left. I_0 \Big( \frac{\sqrt{k_- k_+}}{u} \sqrt{(u t)^2-x^2} \Big) 
	      \theta(t)\, \theta \left(u t - \left| x \right| \right)\right] \ \ .\label{exact2}
\end{eqnarray}
Here $\theta(x)$ denotes the Heaviside step-function and $I_0(x)$ is the modified Bessel function of order zero. 
\begin{figure}[t]
    \includegraphics[trim = 11.9cm 1.0cm 13.3cm 0cm, clip, width=4.2cm]{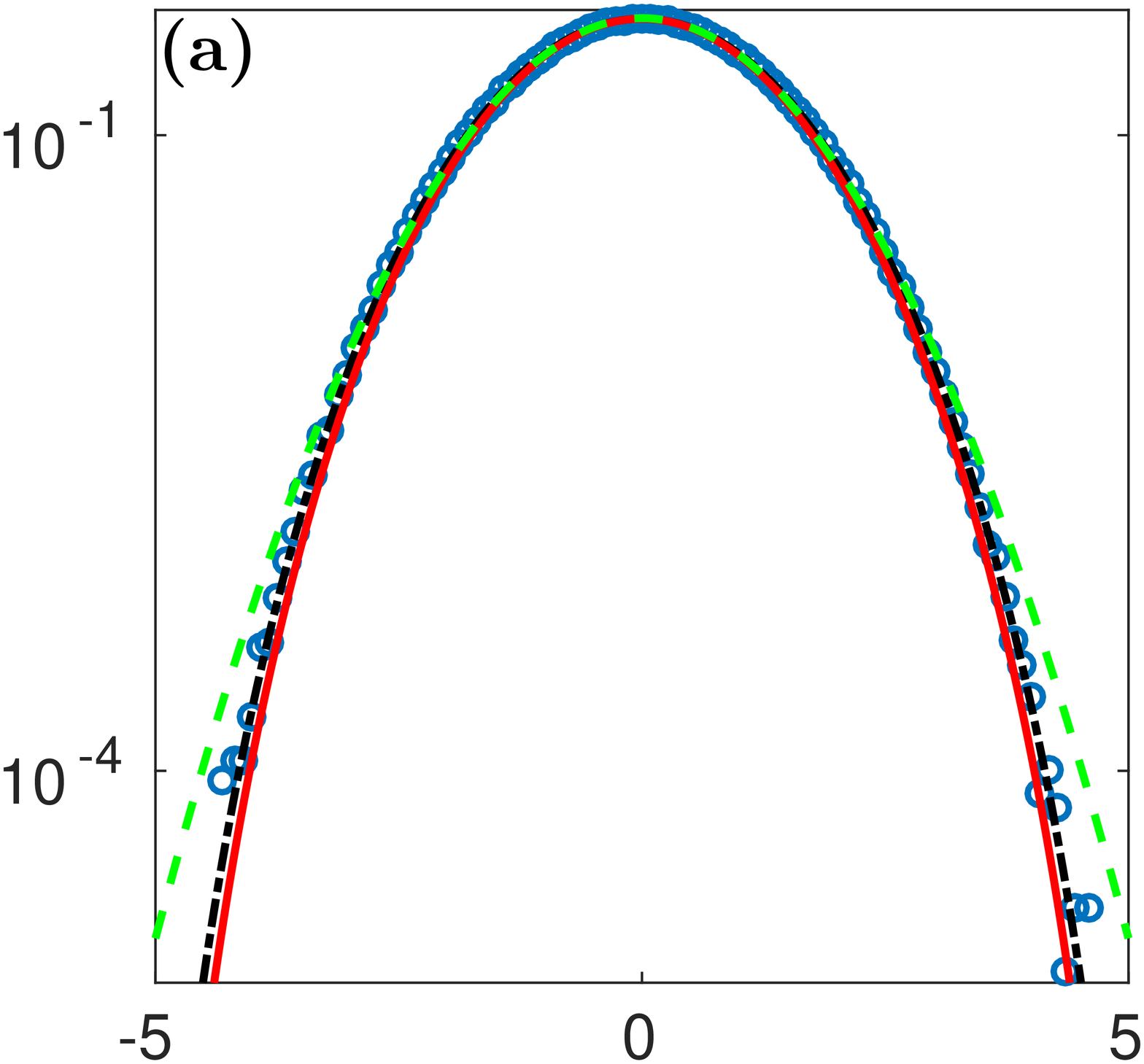}\vspace{-0.1cm}
    \includegraphics[trim = 11.9cm 1.0cm 13.3cm 0cm, clip, width=4.2cm]{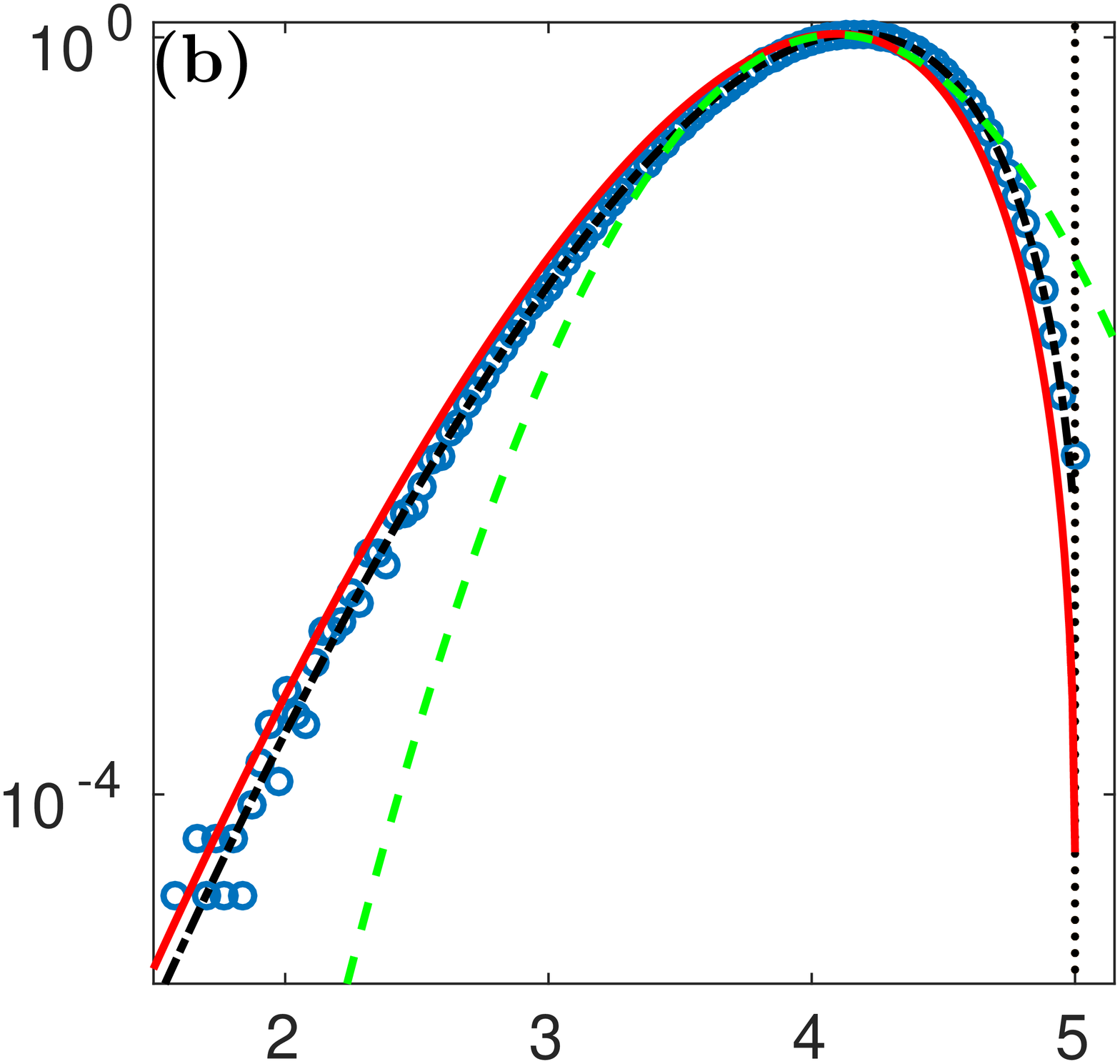}\vspace{-0.1cm}
  \caption{Logarithmic plot of the displacement distribution $P(x,t)$ at $t=10$ for the discrete model of Fig.~\ref{fig:Comparison}b with $N=2$ layers and $u_2=-u_1=.5$. In a) the transition rates are symmetric, $k_+=k_-=2$, in b) they are biased, $k_+ = 10, k_- = 1$. Shown are the exact solution \eqref{exact2} (black), the large deviation result (red), the Gaussian approximation (dashed green), and results from numerical simulations using the Gillespie algorithm for $10^6$ realisation (blue circles). The dotted vertical line in b) markes the cut-off of the distribution at $x=5$.}\label{fig:2-layer model sedimentation t1}
\end{figure}
The ldf $I(v)$ follows directly from \eqref{IJ} because the two constraints $\bar{q}_1+\bar{q}_2=1$ and $u(\bar{q}_1-\bar{q}_2)=v$ already uniquely fix $\bar{\bold{q}}$ in terms of $u$ and $v$. Hence, the contraction becomes trivial:
\begin{equation}\label{eq:2 resIV}
 I(v)=\frac{k_+}{2}(1-\frac{v}{u})+\frac{k_-}{2}(1+\frac{v}{u})-\sqrt{k_+k_-}\sqrt{1-\frac{v^2}{u^2}}. 
\end{equation} 
This result can, of course, also be obtained directly from the large-time limit of \eqref{exact2}.  
In Fig.~\ref{fig:2-layer model sedimentation t1} we exemplarily compare the exact solution \eqref{exact2}, with the Gaussian approximation \eqref{cltI}, and the large-deviation result \eqref{eq:2 resIV}. For the latter we have determined the prefactor such that it coincides with the exact result for $v=\bar{u}$. Included are also results of numerical simulations using the Gillespie algorithm. As expected the large-deviation expression is superior to the Gaussian approximation for reproducing the tails of the distribution and follows the exact result down to rather low probabilities. Note that this is particularly striking for unbalanced transition rates as shown in Fig.~\ref{fig:2-layer model sedimentation t1}b.

\begin{figure}[t]
    \includegraphics[trim = 11.9cm 1.0cm 13.3cm 0cm, clip, width=4.2cm]{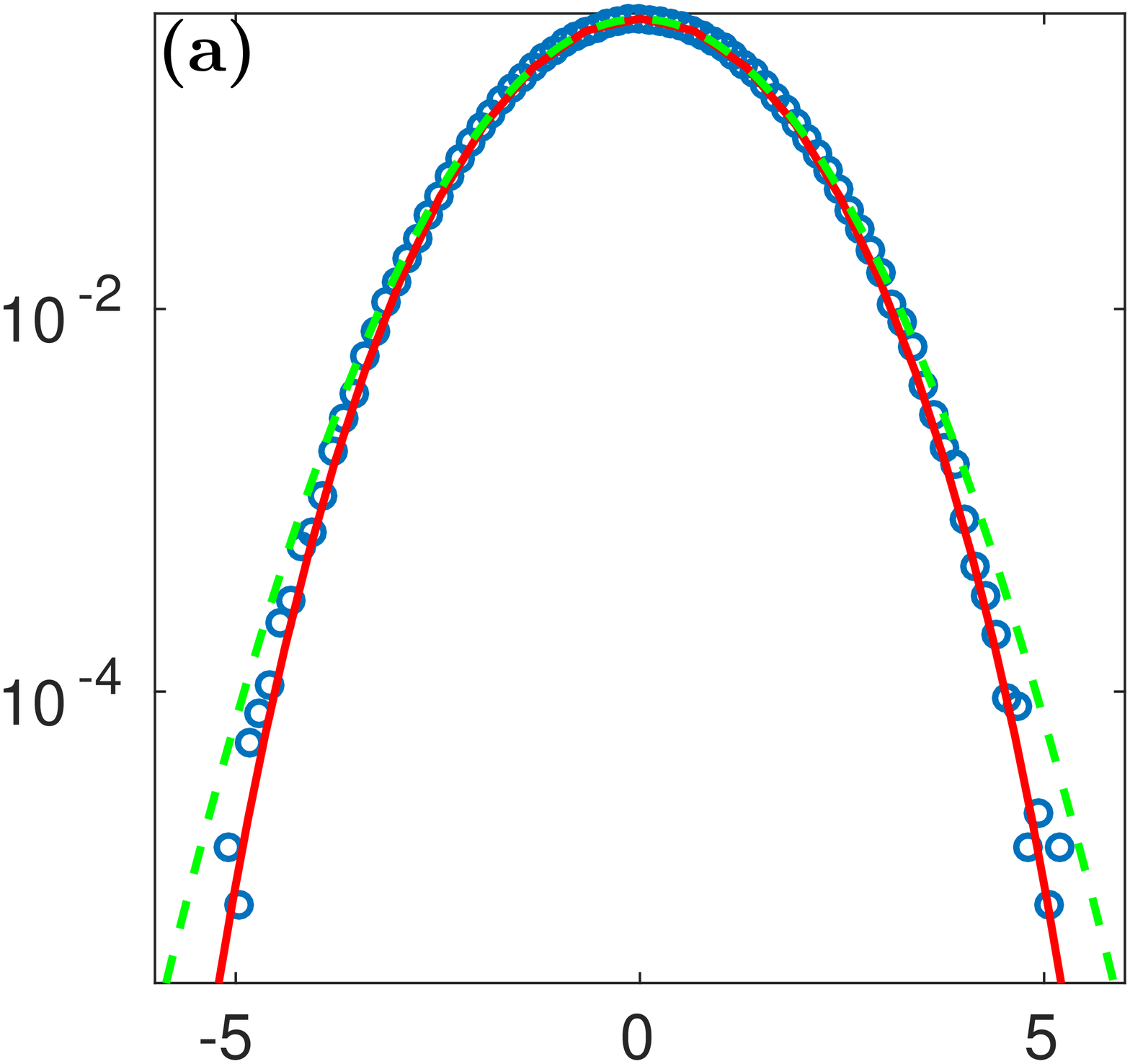}\vspace{-0.1cm}
    \includegraphics[trim = 11.9cm 1.0cm 13.3cm 0cm, clip, width=4.2cm]{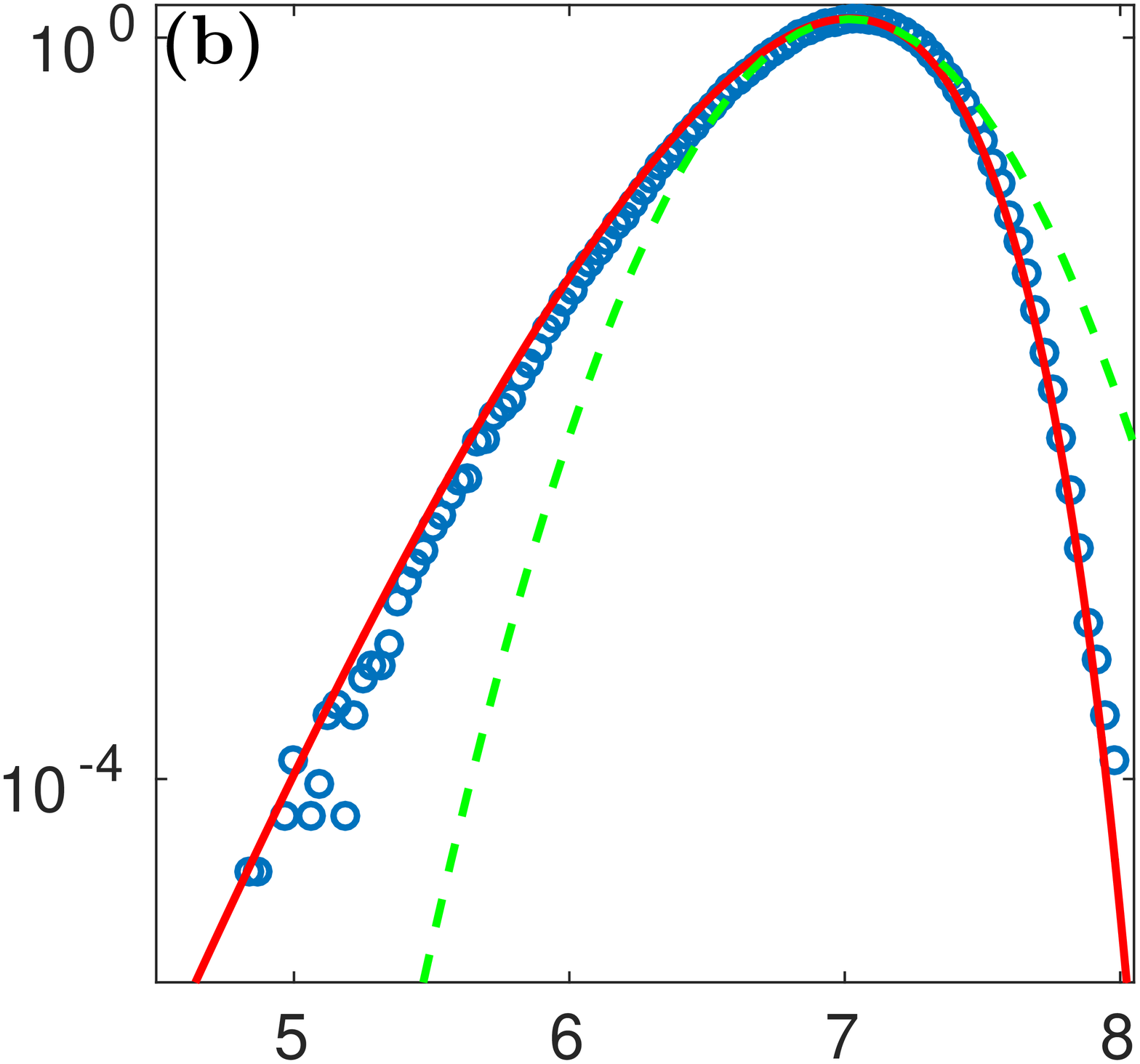}\vspace{-0.1cm}
  \caption{Same as figure \ref{fig:2-layer model sedimentation t1} for a system of $N=10$ layers. The symmetric situation a) is specified by $k_+=k_-=45$ the asymmetric one by $k_+=80, k_-=40$. No exact solution is available for this case.}  \label{fig:N-layer model sedimentation t1}
\end{figure}

We next turn to a discrete model with $N$ layers and nearest neighbour transitions, as sketched in Fig.~\ref{fig:Comparison}b. No explicit analytic result for $I(v)$ is available for the case of general $N$, however, using standard numerical techniques the non-trivial solution $\sqrt{\bvq(\la_1,\la_2)}$ of the homogeneous linear system \eqref{eq:contr} may be determined. Plugging the result into the two constraints results in two equations for $\la_1$ and $\la_2$ from which $I(v)$ is determined via \eqref{Isimple}. For the purpose of illustration, we focus on the simple case of constant upward and downward rates given by $W_{i+1,i}= k_+$ and $W_{i-1,i}=k_-$ for all $i$ (with however boundary rates $W_{N+1,N}=W_{1,0}=0$ ), and consider the simple linear shear flow profile
\begin{equation}\label{eq:linu}
 u_i = - 1 + \frac{2}{N} \left(i - \frac{1}{2} \right)\, .
\end{equation} 
Fig.~\ref{fig:N-layer model sedimentation t1} demonstrates the accuracy of the large deviation approach also for this system. As in Fig.~\ref{fig:2-layer model sedimentation t1} the ldf compares very well with the simulation results and is clearly superior to the Gaussian approximation. The failure of the latter is again particularly prominent for asymmetric transition rates $k_-\neq k_+$. This situation models a horizontal layered system in which Taylor dispersion is affected by sedimentation. Furthermore, the non-Gaussian deviations become stronger when the number of layers $N$ increases. This can be explained by the fact that particles have more options to get stuck in slow or fast lanes. 


\section{Continuous flow fields}\label{sec:cont}
We now return to the case of a continuous variable~$\bold{y}$ instead of the discrete index $i$. If $\bold{y}(t)$ represents a stochastic process with detailed balance we may again build on explicit expressions for the ldf of the empirical density. More precisely, let the dynamics of $\bold{y}$ be given by the Stratonovich Langevin equation 
\begin{eqnarray}\label{eq:LE}
     \partial_t \bold{y}(t) = - \nabla U(\bold{y}) + \sqrt{2D(\bold{y})} \boldsymbol{\xi}(t) 
\end{eqnarray}
with a deterministic drift derived from a potential $U(\bold{y})$ and a Gaussian noise source with zero mean and correlations $\langle \xi_i(t) \xi_j(t') \rangle = \delta_{i,j}\delta(t - t')$. 
The large deviation functional for the empirical density (cf.~\eqref{defq})
\begin{equation}\label{eq:defrho}
 q(\bold{y}):=\frac{1}{t}\int_0^t \delta(\bold{y}-\bold{y}(\tau)) d\tau
\end{equation} 
is then given by \cite{Donskerrev1}:
\begin{equation}\label{Jcont2}
 J[q(\cdot)]=\int  D(\bold{y})\, p(\bold{y})
         \left(\nabla\sqrt{\frac{q(\bold{y})}{p(\bold{y})}}\right)^2 d\bold{y},
\end{equation} 
where, as before, $p(\bold{y})$ denotes the stationary distribution of the process \eqref{eq:LE}. In appendix \ref{sec:B} we include a simple derivation of this expression from \eqref{lded} for the special case of a biased random walk in one dimension. 

Starting with \eqref{Jcont2} the contraction to $I(v)$ is now defined by the variational problem 
\begin{equation}
 I(v)=\inf_{q(\cdot)} \int p(\by)\, D(\by) \left(\nabla \sqrt{\frac{q(\by)}{p(\by)}}\right)^2 d\by
\end{equation} 
under the constraints
\begin{equation}\label{eq:constraints_cont}
 \int q(\bold{y})\,d\bold{y}=1\quad\mathrm{and}\quad \int q(\bold{y}) u(\bold{y})\,d\bold{y}=v\, .
\end{equation} 
It is convenient to minimize in $\hq(\by):=\sqrt{q(\by)}$. Introducing as before the Lagrange multipliers $\la_1$ and $\la_2$ the corresponding Euler-Lagrange equation reads
\begin{equation}\label{ELEgen}
 0=-\frac{1}{\sqrt{p}}\nabla \left(p\,D\,\nabla\Big(\frac{\hq}{\sqrt{p}}\Big)\right)+(\la_1+\la_2 u)\,\hq
\end{equation} 
where at the boundary the normal component of 
\begin{equation}\label{ELEbc}
 \sqrt{p}\,D\,\nabla\big(\frac{\hq}{\sqrt{p}}\big)
\end{equation} 
has to vanish. Multiplying \eqref{ELEgen} by $\hq$, integrating over $\by$, and using the boundary condition \eqref{ELEbc} gives back \eqref{Isimple}. Eqs.~\eqref{ELEgen} and \eqref{ELEbc} form a Sturm-Liouville problem, its concrete solution depends on the flow profile $u(\by)$.

Note that for $D(y)$ independent of $y$, which is the case in many physical applications, the large deviation function becomes proportional to $D$. For the original problem of Taylor dispersion in a Poiseuille flow \eqref{uofr} through a cylindrical tube one can, by a choice of the units of time and space, furthermore set $a=1$ and $\bar{u}=1$. Hence the corresponding large deviation function $I(v)$ divided by $D$ reduces to a single master function, which we proceed to calculate below.

Starting with $p(\by)=$const. and $u(\by)=2(1-\by^2)$ \eqref{ELEgen} acquires the form 
\begin{equation}
 0=-\Delta \hq +\Big(\la_1+2\la_2\,(1-\by^2)\Big)\,\hq.
\end{equation} 
Due to rotational symmetry the solution of this equation will depend only on $r=|\by|$ and it is sufficient to keep the radial part of the Laplacian. We hence have to solve
\begin{equation}\label{eq:ELEsup}
 0=\partial_r^2 \hq(r)+\frac{1}{r}\partial_r \hq(r)-(\lambda_1+2\lambda_2 \, (1-r^2))\,\hq
\end{equation}
together with the boundary conditions
\begin{equation}\label{eq:bcELEsup}
 r\, \partial_r \hq(r)|_{r=0}=0 \qquad\mathrm{and}\qquad \partial_r \hq(r)|_{r=1}=0 .
\end{equation} 
This eigenvalue problem was also obtained in \cite{HaVa} (their Eq.~(3.28)). The auxiliary function $\Phi(r)$ introduced in this paper is hence nothing but the square root $\hq(r)$ of the empirical density. 

The general solution of \eqref{eq:ELEsup} may be written as 
\begin{eqnarray}\label{kummer}
    \hq(r) = &&C_1\, e^{-\frac{\sqrt{B}r^2}{2}}\, M\left(A, 1, \sqrt{B} r^2 \right) + \nonumber\\ 
             &&C_2\, e^{-\frac{\sqrt{B}r^2}{2}}\, U\left(A, 1, \sqrt{B} r^2 \right)
\end{eqnarray}
with the Kummer functions $M(a,b,z)$ and $U(a,b,z)$ \cite{AS}, two constants $C_1$ and $C_2$ as well as 
\begin{equation}
 A=\frac{1}{2}\left(1-\frac{i}{2}\frac{\la_1+2\la_2}{\sqrt{2\la_2}}\right)
\quad\mathrm{and}\quad    B=-2\la_2.
\end{equation} 
The first part of \eqref{kummer} involving $M(a,b,z)$ obeys the first boundary condition, at $r=0$, automatically. Because of the behaviour of $U(a,1,z)$ for small $z$ this boundary condition cannot be fulfilled by the second part of \eqref{kummer} implying $C_2 = 0$. The second boundary condition, at $r=1$, and the two constraints \eqref{eq:constraints_cont} then form a system of equations to determine $C_1$, $\lambda_1$, and $\lambda_2$. From the latter two $I(v)/D$ follows using \eqref{Isimple}.

\begin{figure}[t]
    \includegraphics[trim = 11.9cm 1.0cm 13.3cm 0cm, clip, width=4.2cm]{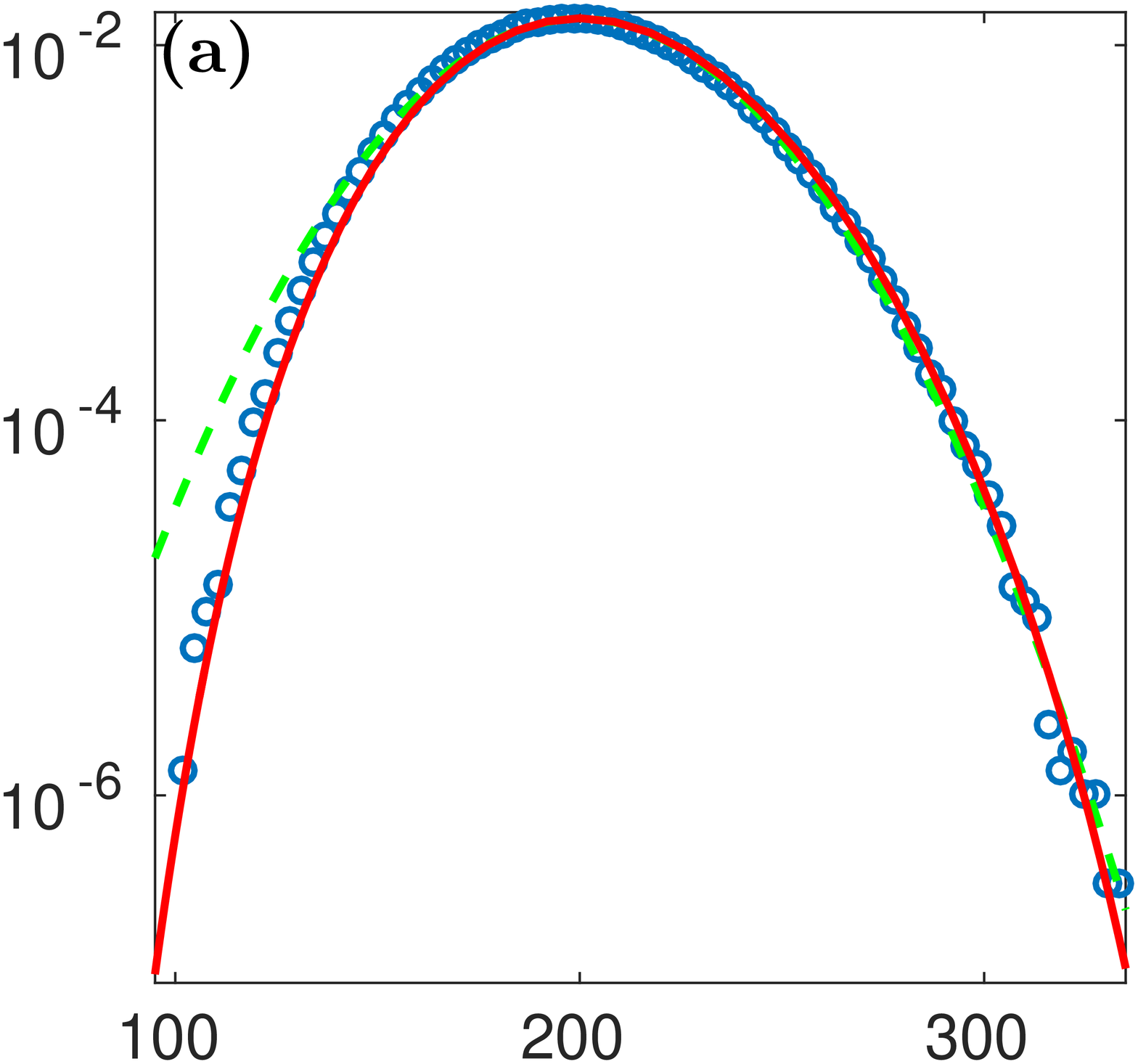}\vspace{-0.1cm}
    \includegraphics[trim = 12.0cm 1.6cm 14.0cm 0cm, clip, width=4.2cm]{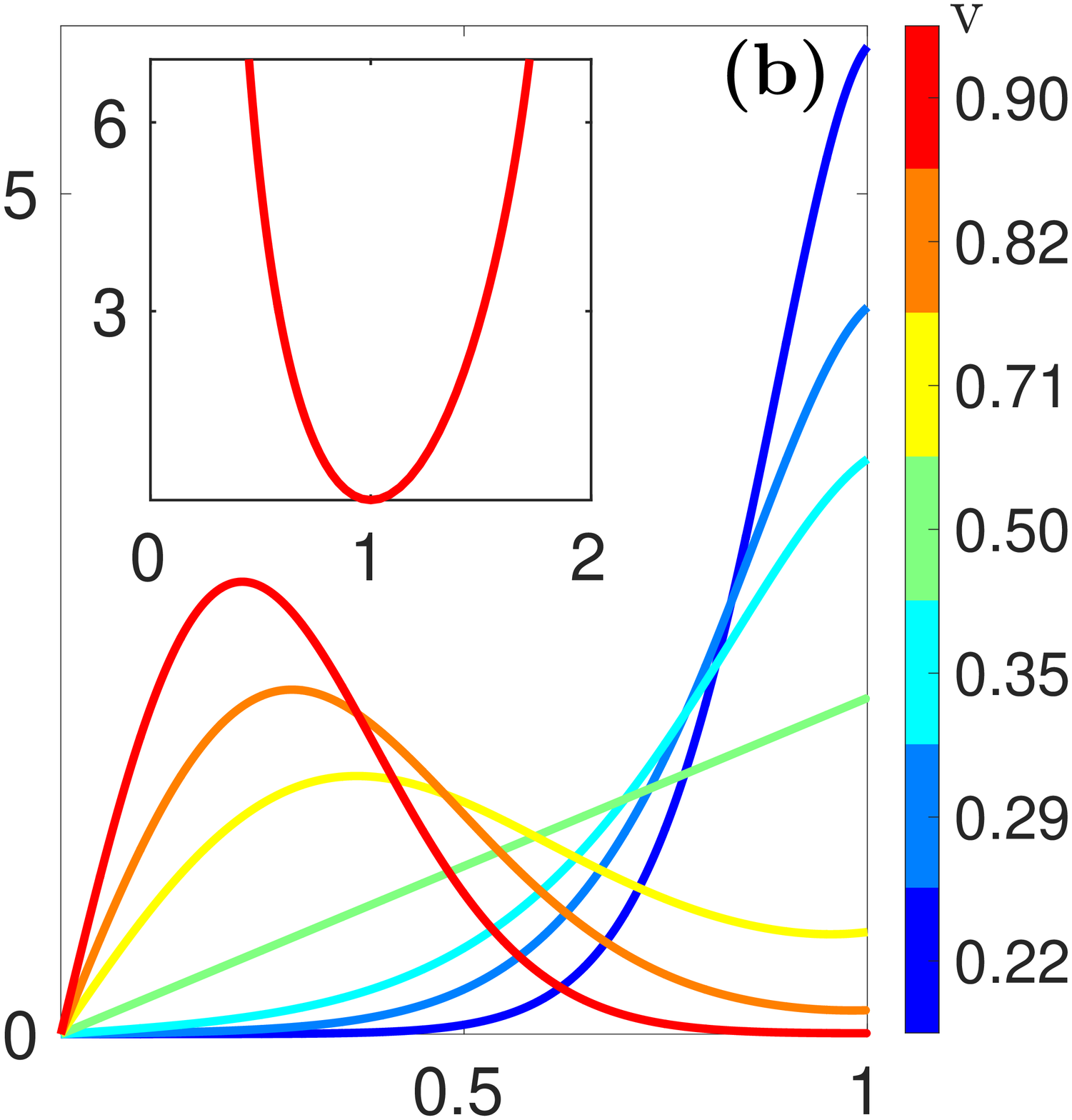}\vspace{-0.1cm}
  \caption{a) Logarithmic plot of the displacement distribution $P(x,t)$ at $t=200$ for a Poisseuille flow \eqref{uofr} with $a=1$ and $\bu=1$ and $D=0.01$. Shown are the large deviation result (red), the Gaussian approximation (dashed green), and the outcome of numerical simulations (blue circles). b) Plot of $\bq(r)$ for different values of $v$ identifying regions of the tube cross section that dominate the various values of $v$. The inset displays the function $I(v)/D$ that universally characterizes the large deviation properties of Taylor dispersion in Poisseuille flow.}\label{fig:Poiseuille}
\end{figure}

In Fig.~\ref{fig:Poiseuille}a we show the ldf obtained in this way together with the Gaussian approximation and numerical simulations. Again, the ldf compares very well with the numerical results. In particular the asymmetry of the distribution, which in this case is due to the circular geometry of the setup, and which is clearly beyond the capabilities of the Gaussian approximation is accurately reproduced. Fig.~\ref{fig:Poiseuille}b displays the function $\bar{q}(r)$ for different values of $v$ that highlights which regions in the cross section of the pipe contribute most to the respective large deviations of the sample velocity. As intuitive small velocities require long residence of the particles near the wall whereas large values of $v$ are realized by particles staying most of the time near the centre of the tube. The inset shows the universal function $I(v)/D$ for Taylor dispersion in Poiseuille flow. Note that both curves are independent of $t$. 


\section{Discussion}\label{sec:discussion}
We discussed the connection between the large deviation properties of Taylor dispersion and the empirical measure of the underlying stochastic process. The large deviation function $I(v)$ provides an accurate long-time characterization of particle separation in shear flows. In particular for biased transport perpendicular to the flow direction the Gaussian approximation was found to yield a rather poor description only whereas the large deviation function reproduces the actual distribution down to very small probabilities. We conclude with a discussion about a converse issue, namely the interest and relevance of Taylor dispersion for the theory of large deviations. First,  results from Taylor dispersion can feedback into the theory of empirical distributions. An example is the exact solution for the dispersion in a general nearest neighbour random walk \cite{BroeckMazo}, which can be translated back into an exact expression for the correlations of the empirical distribution. Second, we note that the ldf $I(v)$, albeit obtained via contraction from the ldf $J({\bf q})$, is in fact a functional of the velocity field $\bold{u}$. Hence, if we know $I_\bold{u}(v)$ for a suitable chosen set of flow profiles $\bold{u}$ we can in principle reconstruct $J(\bold{q})$. Third, as far as experiments are concerned, Taylor dispersion provides a rather direct way to measure the empirical distribution. If the transition matrix governing the $\by$-dynamics can be modified experimentally, e.g., by applying appropriate biasing fields between the states, the results of this "tilt" will be promptly seen in the particle dispersion along the flow.

\appendix

\section{Expressions and bounds for the Taylor dispersion coefficient $K$} \label{sec:A}
Here we give a short account on how our results for the large deviation function $I(v)$ of Taylor dispersion relates to results for the Taylor dispersion coefficient $K$ derived previously. Since $K$ describes the Gaussian long time regime of the process it is determined by the second derivative of $I(v)$ at $v=\bu$:
\begin{equation}\label{KfromJ}
 \frac{1}{2K}=\frac{d^2 I}{dv^2}(\bu)=\frac{d^2 J}{d v^2}(\vep),
\end{equation} 
where $J(\vq)$ is given by (cf.~\eqref{lded}) 
\begin{equation}\label{defJ}
 J(\vq)=\frac{1}{2}\sum_{i,j}\left(\sqrt{W_{i,j}q_j}-\sqrt{W_{j,i}q_i}\right)^2,
\end{equation} 
and $\bvq(v)$ is determined by
\begin{equation}\label{defbarq}
 \frac{\pa J}{\pa q_i}(\bvq)=-(\la_1+\la_2 u_i).
\end{equation} 
Then, since $I(v)=J(\bvq(v))$,
\begin{equation}\label{diffJq}
 \frac{d^2 I}{dv^2}=\frac{d^2 J(\bvq)}{dv^2}=\sum_{i,j} \frac{\pa^2 J}{\pa q_i\pa q_j}(\bvq)
        \,\frac{d \bq_i}{d v}\frac{d \bq_j}{d v}
     +\sum_i \frac{\pa J}{\pa q_i}(\bvq)\,\frac{d^2 \bq_i}{d v^2}.
\end{equation} 
From the constraints  $\bold{1}\cdot\bvq=1$ and $\vu\cdot\bvq=v$ we find by differentiation with resprect to $v$ 
\begin{equation}\label{diffconstraints}
 \bold{1}\cdot \frac{d \bvq}{d v}=0,\quad \vu\cdot\frac{d \bvq}{d v}=1, \quad
 \bold{1}\cdot \frac{d^2 \bvq}{d v^2}=\vu\cdot\frac{d^2 \bvq}{d^2 v}=0.
\end{equation} 
Hence, together with \eqref{defbarq}, it follows 
\begin{equation}
 \sum_i \frac{\pa J}{\pa q_i}(\bvq)\,\frac{d^2 \bq_i}{d v^2}=-\sum_i (\la_1+\la_2 u_i)\,\frac{d^2 \bq_i}{d v^2}=0
\end{equation} 
and the second term in \eqref{diffJq} vanishes. Therefore
\begin{equation}\label{res1A}
 \frac{1}{2K}=\sum_{i,j} \frac{\pa^2 J}{\pa q_i\pa q_j}(\vp)
        \,\frac{d \bq_i}{d v}(\bu)\frac{d \bq_j}{d v}(\bu).
\end{equation} 
Using the detailed balance condition $W_{i,j}p_j=W_{j,i}p_i$ we find from \eqref{defJ}
\begin{equation}\label{h2}
  \frac{\pa^2 J}{\pa q_i\pa q_j}(\vp)=-\frac{1}{2}\frac{W_{i,j}}{p_i}=-\frac{1}{2}\frac{W_{j,i}}{p_j},
\end{equation} 
and therefore
\begin{equation}\label{res2A}
\frac{1}{K} =-\sum_{i,j} \frac{W_{i,j}}{p_i}\,\frac{d \bq_i}{d v}(\bu)\frac{d \bq_j}{d v}(\bu).
\end{equation} 

To determine $d \bq_i/dv$ at $v=\bu$ we take the derivative of \eqref{defbarq} with respect to $v$ to find 
\begin{equation}\label{h1}
 \sum_j \frac{\pa^2 J}{\pa q_i\pa q_j}(\bvq)\,\frac{\pa \bq_j}{d v}=-\frac{d \la_1}{d v}-\frac{d \la_2}{d v} u_i, 
 \qquad i\in\bold{I}.
\end{equation} 
Since $J$ is a homogeneous function of $\vq$, $J(\kappa \vq)=\kappa J(\vq)$, we have 
\begin{equation}
 J(\vq)=\sum_i \frac{\pa J}{\pa q_i} q_i
\end{equation} 
and, for $\vq=\bvq$, 
\begin{equation}\label{Jofla}
 J(\bvq)=-\sum_i (\la_1+\la_2 u_i) \bq_i=-\la_1-\la_2 v,
\end{equation} 
reproducing \eqref{Isimple}. This implies 
\begin{equation}
 \frac{d J(\bvq)}{d v}=-\frac{d \la_1}{d v}-\frac{d \la_2}{d v} v -\la_2.
\end{equation} 
On the other hand, multiplying \eqref{defbarq} by $\partial \bq_i/\pa v$ and summing over $i$ gives
\begin{equation}\label{diffJv}
\frac{d J(\bvq)}{d v}=-\la_2(v) 
\end{equation} 
such that 
\begin{equation}
 \frac{d \la_1}{d v}=-\frac{d \la_2}{d v} v.
\end{equation} 
Therefore
\begin{equation}\label{diffla2}
 -\frac{d \la_2}{d v}(\bu)=\frac{1}{2K}\qquad\mathrm{and}\qquad\frac{d \la_1}{d v}(\bu)= \frac{1}{2K}\bu.
\end{equation} 
From \eqref{h1} we hence obtain
\begin{equation}
 \sum_j \frac{\pa^2 J}{\pa q_i\pa q_j}(\vp)\frac{d \bq_j}{d v}=\frac{\bu-u_i}{2K}.
\end{equation} 
or, using \eqref{h2},
\begin{equation}\label{linsys}
 \sum_j W_{i,j}\frac{d\bq_j}{d v}(\bu)=p_i\frac{\bu-u_i}{K}.
\end{equation} 
This is a system of linear inhomogeneous equations to determine $d \bq_i/dv$ at $v=\bu$ in terms of $\vp$ and $\vu$.
The matrix $W_{i,j}$ is singular, however, by definition of $\bu$, the r.h.s of \eqref{linsys} is orthogonal to the left eigenvector $(1,1,...,1)$ corresponding to the zero eigenvalue and the system is hence solvable. We denote by $\tilde{W}^{-1}_{i,j}$ the inverse of $W_{i,j}$ on the subspace orthogonal to the zero eigenvector to get
\begin{equation}
 \frac{d\bq_i}{d v}(\bu)=\frac{1}{2K}\sum_k \tilde{W}^{-1}_{i,k}\,p_k\,(\bu-u_k).
\end{equation} 
Plugging this result into \eqref{res2A} we find the well-known expression \cite{Bouten}
\begin{equation}\label{resK3}
 K=-\sum_{i,j} \tilde{W}^{-1}_{i,j}\, p_j\,(\bu-u_i)(\bu-u_j).
\end{equation} 

In addition to this exact expression for $K$ the variational form of the contraction \eqref{IJ} forms a convenient starting point to derive bounds on $K$. From \eqref{resK3} we see that a Galilei transformation $u_i\to u_i+u$ does not change $K$. In the following we therefore assume for simplicity $\bu=0$. Small deviations $v$ from $\bu=0$ result from small deviations of $\bvq$ from $\vp$. If we parametrize these deviations by $q_i=p_i(1+\eps_i v)$ and use the detailed balance condition \eqref{db} we find 
\begin{equation}
 J(\bvq)\cong\frac{v^2}{8}\sum_{i,j} W_{i,j} p_j(\eps_i-\eps_j)^2 ,
\end{equation} 
which, using \eqref{KfromJ} and \eqref{IJ}, yields
\begin{equation}\label{boundK}
 \frac{1}{K}=\inf_{\makebox{\boldmath{$\eps$}}}\; \frac{1}{2} \sum_{i,j} W_{i,j} p_j(\eps_i-\eps_j)^2 ,
\end{equation} 
Here the infimum has to be taken under the constraints $\sum_i\eps_i q_i=0$ and $\sum_i\eps_i q_i u_i=1$. 

As an interesting trial ansatz, we set $\eps_i=\al u_i$. Then the first constraint is automatically satisfied while the second implies 
\begin{equation}
 \frac{1}{\al}=\sum_i u_i^2 p_i=\overline{u^2}.
\end{equation} 
This gives 
\begin{equation}
 \frac{1}{K}\leq \frac{1}{2\overline{u^2}^2} \sum_{i,j} W_{i,j} p_j(u_i-u_j)^2 .
\end{equation} 
The equality sign is realized for a Kangaroo process defined by $W_{i,j}=p_i/\tau$, where $\tau$ denotes the overall relaxiation time. Indeed one finds in this case that the infimum in \eqref{boundK} is realized for $\eps_i=u_i/\overline{u^2}$. Hence for a Kangaroo process with $\bu=0$ we get the simple result
\begin{equation}
 K=\frac{2\tau \overline{u^2}^2}{\sum_{i,j} p_i p_j (u_i-u_j)^2}=\tau\overline{u^2}.
\end{equation} 
Finally, using $\sum_i W_{i,j}=0$ we may rewrite \eqref{boundK} in the form
\begin{align}\nonumber
 \frac{1}{K}&=-\inf_{\makebox{\boldmath{$\eps$}}}\; \frac{1}{2} \sum_{i,j} W_{i,j} p_j \eps_i\eps_j\\\nonumber
            &=-\inf_{\makebox{\boldmath{$\eps$}}}\; \frac{1}{2} \sum_{i,j} \eps_i p_i W_{i,j} \eps_j p_j/p_i\\
            &=-\inf_\vf \lan \vf|W|\vf\ran,\label{boundK2}
\end{align}
where the infimum is now over all vectors $\vf$ orthogonal to the stationary distribution $\lan \vf|\vp\ran=0$ and fulfilling 
$\lan \vf|\vu\vp\ran=1$. Here, the scalar product is defined by 
\begin{equation}
 \lan \vf|\mathbf{g}\ran=\sum_i\frac{f_i g_i}{p_i}.
\end{equation} 
The bound 
\begin{equation}
 K\geq-\frac{1}{\lan \vf|W|\vf\ran}
\end{equation} 
resulting from \eqref{boundK2} was already derived in \cite{Bouten}.

\section{Continuum limit of the ldf for a biased random walk}\label{sec:B}

Consider a biased nearest neighbour random walk in one dimension with reflecting boundary conditions. The probability $P_i(t)$ of the walker being at site $i\in\bold{I}$ at time $t$ obeys a Master equation 
\begin{equation}
 \pa_t P_i(t)=\sum_j \Big(W_{i,j} P_j(t)-W_{j,i} P_i(t)\Big)
\end{equation} 
with 
\begin{equation}
 W_{i,j}=\left\{\begin{array}{cll}
                k_{i-1}^+ &\mathrm{for} &j=i-1\\
                -(k_i^++k_i^-) &\mathrm{for} &j=i\\
                k_{i+1}^- &\mathrm{for} &j=i+1\\
                0 &\mathrm{else} &
               \end{array} \right.
\end{equation} 
The stationary distribution $p_i$ fulfills detailed balance $k_i^+p_i=k_{i+1}^-p_{i+1}$ automatically. Using this condition we find from \eqref{defJ}
\begin{equation}
 J(\vq)=\sum_i k_{i-1}^+ p_{i-1} \left(\sqrt{\frac{q_{i-1}}{p_{i-1}}}-\sqrt{\frac{q_i}{p_i}}\right)^2.
\end{equation} 
The continuum limit is obtained from this expression by introducing a small space interval $\eta$ and using the replacements
\begin{eqnarray}
 i\to\frac{y}{\eta}, \quad p_i\to\eta &&p(y), \quad q_i\to\eta q(y) \nonumber\\ 
 \sum_i\to\int\frac{dy}{\eta}, &&\quad J(\vq)\to J[q(\cdot)] \nonumber\\
 k_i^\pm\to \pm&&\frac{w(y)}{2\eta}+\frac{D(y)}{\eta^2} .
\end{eqnarray} 
The probability density function $P(y,t)$ then fulfills the Fokker-Planck equation 
\begin{equation}
 \pa_t P(y,t)=-\pa_y\Big(\big(w(y)-\pa_y D(y)\big)P(y,t)\Big) + \pa_y\Big(D(y)\pa_y P(y,t)\Big)
\end{equation} 
complemented by reflecting boundary conditions. The large deviation functional acquires the form 
\begin{equation}
 J[q(\cdot)]=\int p(y) D(y) \left(\pa_y \sqrt{\frac{q(y)}{p(y)}}\right)^2 dy
\end{equation} 
corresponding to \eqref{Jcont2}.

\end{document}